\documentclass[11pt]{article}
\usepackage{geometry}                
\usepackage{graphicx}
\usepackage{authblk}

\title{Data Set From Molisan Regional Seismic Network  Events}
\author{Giovanni De Gasperis, Christian Del Pinto}

\affil{Dipartimento di Ingegneria e Scienze dell'Informazione e Matematica,
\\ Universit\'a degli Studi dell'Aquila, Italy}

\begin{document}
\maketitle

\begin{abstract}
After the earthquake occurred in Molise (Central Italy) on 31st October 2002 (Ml 5.4, 29 people dead), the local {\em Servizio Regionale per la Protezione Civile} to ensure a better analysis of local seismic data, through a convention with the {\em Istituto Nazionale di Geofisica e Vulcanologia} (INGV), promoted the design of the Regional Seismic Network (RMSM) and funded its implementation. The 5 stations of RMSM worked since 2007 to 2013 collecting a large amount of seismic data and giving an important contribution to the study of seismic sources present in the region and the surrounding territory. This work reports about the dataset containing all triggers collected by RMSM since July 2007 to March 2009, including actual seismic events; among them, all earthquakes events recorded in coincidence to {\em Rete Sismica Nazionale Centralizzata} (RSNC) of INGV have been marked with S and P arrival timestamps. Every trigger has been associated to a spectrogram defined into a recorded time vs. frequency domain. The main aim of this structured dataset is to be used for further analysis with data mining and machine learning techniques on image patterns associated to the waveforms.
\end{abstract}

\section{Introduction and history of the Molisan Regional Seismic Network}

This work aims to provide to the research community a structured data set of more than two years of seismic events of measurements, so it can be adopted by data mining researchers for more investigations. We present a method of dataset creation  that we adopted to assemble a fully representative set of seismic data coming from the regional seismic network of Molise in Italy, in the years from 2007 to 2009. 

On the 31st of October, 2002, at 11:32 (local time), an earthquake with magnitude Ml = 5.4 occurred in the eastern Molise, Central Italy, near Puglia boundary, interesting a wide area including the territories of San Giuliano di Puglia, Santa Croce di Magliano and Larino \cite{2004decanini}, \cite{2003pinto} , \cite{2004valensise}, \cite{2004macroseismic}. The seismic event caused serious damages to several buildings: the severest one interested the school of San Giuliano di Puglia, that crashed killing 27 children and their teacher. In the same time and in the same place, an old women died for an infarction caused by the fear. The main event had a remarkable aftershock the day after, at 16:08 (local time) with magnitude $M_l = 5.3$, at about 12 km in SW direction from the first epicenter.
In the second half of 2006 , one of the author, employed at that time as scientific advisor of molisan {\em Centro Funzionale del Servizio Regionale per la Protezione Civile} , proposed the design of a regional seismic network; starting requirements were to be compatible with the national seismic network \cite{1998analisi} and to be maintained by a public authority. Then the local Government decided to improve the \emph{Servizio Regionale per la Protezione Civile (Regional Service for Civil Protection)}, giving a relevant priority to seismic studies and monitoring. The project was then officially initiated by a specific legal deliberation \cite{burm2006} and by the signature of an agreement between the local \emph{Servizio Regionale per la Protezione Civile} and the \emph{Istituto Nazionale di Geofisica e Vulcanologia (INGV)}. 
First two stations were installed, after a period spent in detections of suitable sites, in July 2007 in the unmanned school of Civita di Bojano and in the graveyard of Sant'Angelo in Grotte; both sites were located inside the Matese mountain range, identified as the most seismogenic area in Molise after a preliminary study on historical and instrumental data \cite{2003ground}, \cite{2003cdp1}, \cite{2003cdp2}, \cite{2005cdp}. All data were sent, in real time, to the Seismic Room established in the premises of the \emph{Centro Funzionale} in the territory of Campochiaro; here the Molise Region installed the same hardware and software setup used in the I.N.G.V.  Seismic Room in Rome in order to have a complete integration with the data from Centralized National Seismic Network (RSNC). The real time data channel comprised a dedicated Hiperlan radio network with 5.6 GHz (dorsal link) and 2.4 GHz (seismoterminal link) frequencies. INGV provided the proprietary platform of all installations, including multi-parametric seismic stations and 1-Hz seismometers. The sampling rate was fixed at 100 Hz.

\begin{figure}
\includegraphics[scale=0.9]{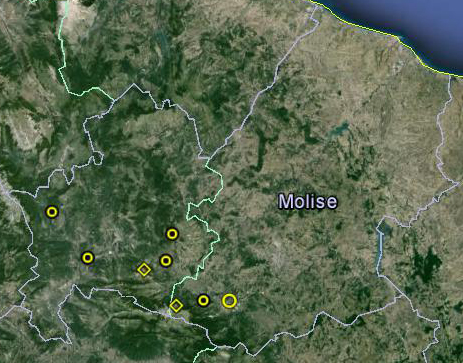}
\caption{Final configuration of Molisan Regional Seismic Network structure (from April 2008 to September 2013). Seismic stations are represented by inner dark circles, seismic room in Centro Funzionale Regionale by empty circle and Hiperlan radio bridges by empty squares. \emph{\small (satellites photo courtesy of Google Earth)} }
\label{rmsm}
\end{figure}

In the April 2008 three new stations were installed, in the locality of Colle dell'Orso near Frosolone, in a museum cellar of Castel San Vincenzo and inside the crypt of the Santa Maria \emph{ad Macle} church near Macchia d'Isernia. By this final structure, comprehending 5 seismic stations, it was also possible to record little local events that were  not recorded by the RSNC by INGV \cite{2008cdp}. The final configuration is shown in Fig. \ref{rmsm} .

The importance of the RMSM may be found not only in collecting data, but also due to the increased granularity it offers to the national network, giving the possibility to provide a more accurate events localization, especially for local earthquakes. Moreover, this upgrade allowed to improve quality and number of low magnitude events, useful to give a contribution to the local seismic hazard maps. These events, in fact, are able to give information related to the seismogenic activity in a certain area, and their data can be used in order to study the characteristics of excitations and propagation of seismic waves in the molisan section of Apennines, allowing in this area a detailed study. The collected data will be also able to estimate its ground motion scaling \cite{2003ground}, \cite{2003cdp1}, \cite{2003cdp2}, \cite{2005cdp}, that describes a higher resolution approximation than considering the hypothesis of crustal homogeneity.
The RMSM network was active until the first half of September 2013. After this date, the \emph{Servizio Regionale per la Protezione Civile} (changed, in the meanwhile, in an agency named \emph{Agenzia Regionale per la Protezione Civile}) has lost interest to maintain the quality of service of the network, causing its progressive disposal which was fully completed in the first months of 2016, after more than two years of data loss. Since July 2007 to September 2013, the Molisan Regional Seismic Network was able to collect a huge data-set which gave an important support to national seismicity researches (as, for example, SLAM, about seismicity from Marche, Abruzzo and Molise regions), offering an increased knowledge on local seismicity comparing to the period when only RSNC was active on that territory.

\section{RMSM Data Description}
Over the RMSM operational period, the collected data is partitioned into three raw subsets:

\begin{enumerate}
\item Period from July 2007 to March 2008: only two seismic stations were active. In these period was possible to count only 9 days of data-loss.
\item Period from April 2008 to April 2010: the number of seismic station was increased to five. This was the longest period of RMSM with high availability, when occurred only 11 days of data-loss. In these 25 months, the RMSM collected most of data from the seismic crisis followed the L'Aquila earthquake1 \cite{2009laquila}, occurred on the 6th of April 2009, which added considerable "noise" in the local events data-set.
\item Period from May 2010 to September 2013: in this period a lower priority management of Hiperlan network, a not-renewed convention with INGV and a progressive loss of local administrator's interests, caused a huge decreasing in number and quality of data collected. The RMSM was mostly offline \cite{2011cdp}.
\end{enumerate}

\begin{figure}
\includegraphics[scale=1.45]{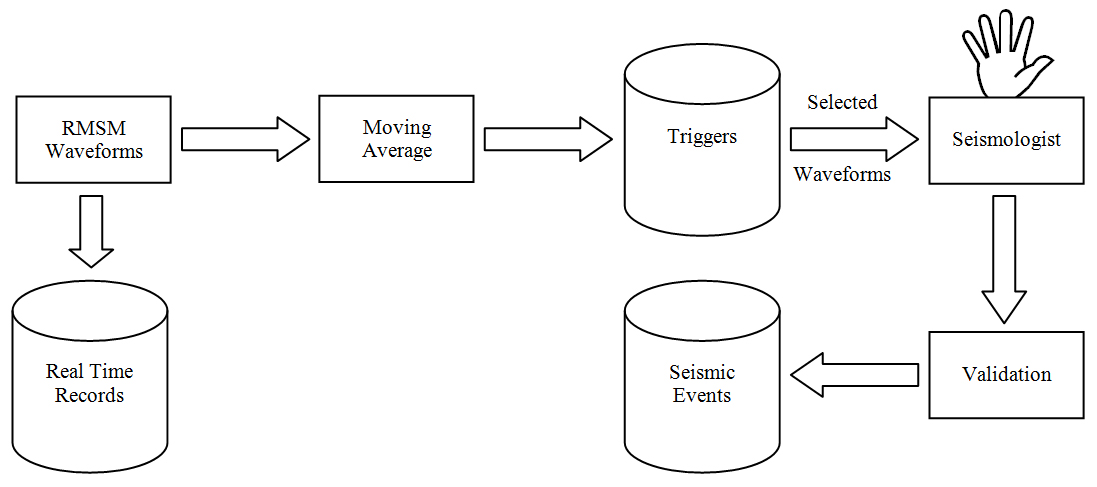}
\caption{Validation process on collected data}
\label{diagram}
\end{figure}

Due to the reasons previously exposed, we decided to use only the seismic data collected in the period July 2007 to March 2009, covering continuous period of 21 months. This period comprises 1350 local seismic events, identified by conventional human guided methods; each event was recorded concurrently by 1 to 5 stations, while only 459 separately collected, in the same period, by the RSNC. 
The acquisition program used to discriminate triggers from real time waveforms applied the common method of moving average \cite{2010sharma}. 
All detected triggers had been later validated to identify (and locate, if possible) seismic events, as shown in Fig.\ref{diagram}.

\section{Data Set}

The 1350 seismic events comprised 6348 waveforms, divided on the three channels (vertical Z, horizontal N-S and horizontal E-We ordered coincidences coming from 1 to 5 stations. We selected the Z channel to be included in the dataset since it contains the most information content to identify a seismic event, i.e. P wave. We then partitioned the coincidences by every measuring station obtaining 2086 seismic waveforms \cite{2007cdp}, among which 946 validated through correspondence events recorder and published by RSNC on the BSI ({\em Bollettino della Sismicit\'a Italiana}) \cite{bsi} released every 6 months by INGV, index by origin time. The total number of triggers with an associated spectrogram is 51087, for a total of 3.62GB of data.

\begin{figure}
\hspace{.2\textwidth}
A: \includegraphics[scale=0.6]{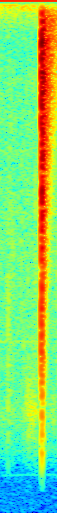}
\hspace{.2\textwidth}
B: \includegraphics[scale=0.6]{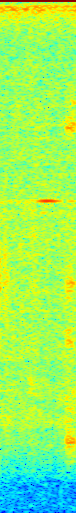}
\caption{Waveforms spectrograms from CVB1 station. 
A:  Seismic waveform referred to a local earthquake occurred in Guardiaregia, Molise, Italy, on 12 January 2008 in the range of 10Km from the station, at lat:41.392 N, long:14.496 E, depth: 5.7Km, origin time: 23:39:48.6, Ml: 2.8 {\em (localization data from INGV BSI)}. B: No seismic waveform.}
\label{spectra}
\end{figure}

The program {\tt MakeSpectra.py} can read the raw waveform by means of the ObsPy Python framework for seismology \cite{2010obspy} and calculate the FFT using the NumPy/MatPlotLib libraries. The FFT parameters are the following: lowpass filter at 25Hz, 2 corners, dt = 0.02 s, num. of FFT samples: 1024. Two examples are shown in Fig. \ref{spectra} , local seismic waveform vs. a generic triggered waveform (non seismic). The horizontal pixel scale represent the time scale, the vertical pixel scale represent the frequency scale. The horizontal size is variable since it depends from the station recording system.

\noindent The final waveform meta-data table has the following fields schema:

\vspace{0.25 cm}
\noindent {\tt Station ID | Year | Month | Day | Start | P arrival | S arrival | Spectrogram }
\vspace{0.25 cm}

where:
\begin{itemize}
\item \textbf{Station ID}:  a 4-characters alphanumeric code needed to identify a particular station;
\item \textbf{Date}:  year, month  and day  when the particular event occurred;
\item \textbf{Start}: start recording timestamp with the format ''{\tt <H>:<M>:<S>}", with integer seconds;
\item \textbf{P arrival}: a timestamp with the format ''{\tt <H>:<M>:<S.xx>}", with 2 decimals for hundredths of seconds related to the P-waves arrival time at a particular station, validated by the seismologist through manual picking procedures; it may be null;
\item \textbf{S arrival}: a timestamp with the format ''{\tt <H>:<M>:<S.xx>}", with 2 decimals for hundredths of seconds related to the S-waves arrival time at a particular station, validated by the seismologist through manual picking procedures; it may be null;
\item \textbf{Spectrogram}: name of the spectrogram file calculated with the {\tt MakeSpectra.py} program from the raw waveform. 
\end{itemize}

We restricted the dataset only to the vertical channel because during the peaking detection this channel allows to better measure the P-waves arrival time. Columns P and S arrival may contains null values for those waveforms were picking was not done, or they were not related to a seismic event.

\begin{figure}
\includegraphics[scale=0.35]{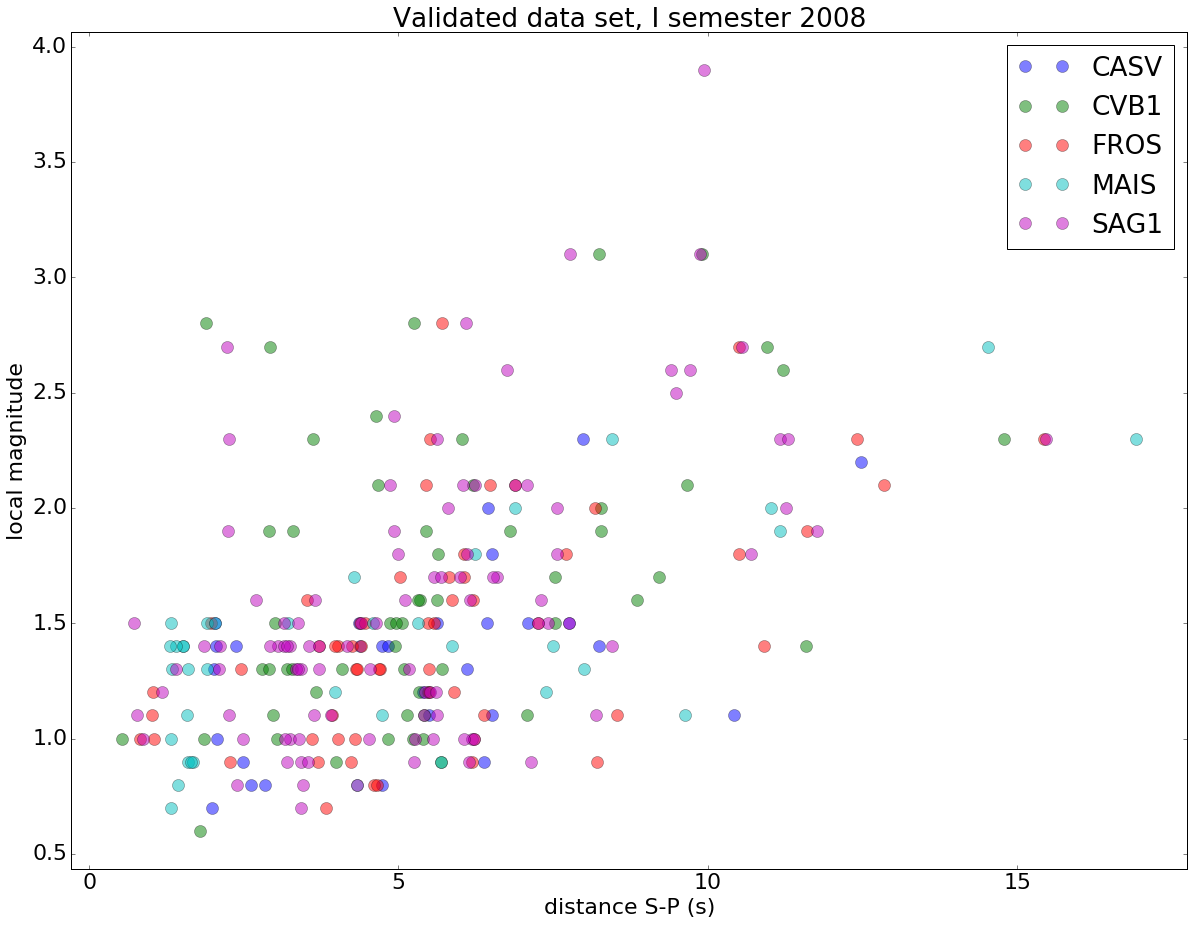}
\caption{Validated data set portion scatter plot, I semester 2008 events from RMSM, grouped by station. The distance S-P arrival times gives an estimate of focus distance from the station. In average events are in the range of about 5s (less than 40Km) and most lower than Ml 2 }
\label{scatter}
\end{figure}

An overview of the I semester 2008 of the data set, with 282 events collected by RMSM and validated by RSNC, is shown in Fig. \ref{scatter}.

The dataset is publicly available at: \cite{dataset} .

\section{Conclusion}

We collected seismic events from the Molise Region in Italy between 2007 and 2009, including certified earthquakes, no certified earthquakes and triggers derived from non seismic events. For each raw trigger we built a spectrogram, i.e. an associated image pattern. 
The data set is structured so to be analyzed with data mining/machine learning techniques to distinguish actual seismic waveforms among raw triggers, without human intervention.

\section{Acknowledgements}

We thank Dr. Gaetano De Luca at INGV for guiding during data collection and insightful discussion. This work has been possible by in internal grant from {\em Dipartimento di Ingegneria e Scienze dell'Informazione e Matematica}, University of L'Aquila. 

\bibliographystyle{IEEEtran}
\bibliography{seismic}

\end{document}